**Denoising method for dynamic contrast-enhanced CT perfusion studies using three-dimensional deep image prior as a simultaneous spatial and temporal regularizer**

Kenya Murase

Department of Future Diagnostic Radiology, Graduate School of Medicine, Osaka University, Suita, Osaka, Japan

Corresponding author:

    Kenya Murase, PhD

    Department of Future Diagnostic Radiology, Graduate School of Medicine,

    Osaka University

    2-2 Yamadaoka, Suita, Osaka 565-0871, Japan

    Tel & Fax: (81)-6-6879-3435

    E-mail: murase@sahs.med.osaka-u.ac.jp

**Abstract**

This study aimed to propose a denoising method for dynamic contrast-enhanced computed tomography (DCE-CT) perfusion studies using a three-dimensional deep image prior (3D DIP), and to investigate its usefulness in comparison with total variation (TV)-based methods with different regularization parameter ($\alpha$) values through simulation studies. In the proposed DIP method, the 3D DIP was incorporated into the constrained optimization problem for image denoising as a simultaneous spatial and temporal regularizer, which was solved using the alternating direction method of multipliers. In the simulation studies, DCE-CT images were generated using a digital brain phantom and their noise level was varied using the X-ray exposure noise model with different exposures (15, 30, 50, 75, and 100 mAs). Cerebral blood flow (CBF) images were generated from the original contrast enhancement (CE) images and those obtained by the DIP and TV methods using block-circulant truncated singular value decomposition. The quality of the CE images was quantitatively evaluated using the peak signal-to-noise ratio (PSNR) and structural similarity index measure (SSIM). To compare the CBF images obtained by the different methods and those generated from the ground truth images, linear regression analysis was performed to calculate the correlation coefficient between them and the slope and y-intercept of the regression equation. When using the DIP method, the PSNR and SSIM were not significantly dependent on the exposure, and the SSIM was the highest for all exposures. When using the TV methods, they were significantly dependent on the exposure and $\alpha$ values. The results of the linear regression analysis suggested that the linearity of the CBF images obtained by the DIP method was superior to those obtained from the original CE images and by the TV methods. Our preliminary results suggest that the DIP method is useful for denoising DCE-CT images at ultra-low to low exposures ($\leq$ ~50 mAs) and for improving the accuracy of the CBF images generated from them.

## 1. Introduction

Dynamic contrast-enhanced computed tomography (DCE-CT) is a promising tool for analyzing haemodynamic changes in tissues and organs (Miles 1991). Cerebral perfusion studies can be performed using DCE-CT and cerebral perfusion parameters, such as cerebral blood flow (CBF) can be quantified. Assessing the state of perfusion in the brain using these perfusion parameters is important for determining medical treatment plans and/or prognostic prediction in patients with cerebrovascular diseases (Wintermark *et al*. 2002).

Radiation exposure during DCE-CT perfusion studies is a serious problem, and is one of the obstacles preventing them from becoming widespread (Hirata *et al*. 2004). Reducing the X-ray tube current of CT is one of the methods used to reduce the radiation dose to patients. However, this increases the statistical nose in CT images (Murase *et al*. 2005). To reduce the noise in medical images, various methods have been devised, including filtering techniques, such as Gaussian filtering (Nagayoshi *et al*. 2005), time-intensity profile similarity bilateral filtering (Mendrik *et al*. 2011), and partial differential equation-based methods, such as an anisotropic diffusion method (Murase *et al*. 2015). With the emergence of compressed sensing (Lustig *et al*. 2007), sparsity-inducing regularizers, such as the total variation (TV) norm have been used as powerful tools for enforcing piece-wise smoothness in image processing and DCE-CT perfusion studies (Beck and Tebouble 2009; Niu *et al*. 2016; Zeng *et al*. 2016; Li *et al*. 2019).

In recent years, denoising methods based on convolutional neural networks (CNNs) have been developed and have achieved significant success (Zhang *et al*. 2017; Chen *et al*. 2017; Kadimesetty *et al*. 2019). When applying these methods to CT images, CNNs are trained to map low-dose CT images to normal- or high-dose CT images using a training set with paired or unpaired low- and normal- or high-dose CT images (Zhang *et al*. 2017; Chen *et al*. 2017). Recently, the use of a deep image prior (DIP) has been proposed (Ulyanov *et al*. 2018), and recent reports have shown that this approach is useful for image denoising, restoration, and reconstruction (Gong *et al*. 2019; Zhou and Horstmeyer 2020; Yoo *et al*. 2021). This approach is based on the finding that CNNs alone have an inherent bias towards natural images and requires no prior training pairs, contrary to the above supervised training approaches (Ulyanov *et al*. 2018).

This study aimed to propose a denoising method for DCE-CT perfusion studies using three-dimensional (3D) DIP, and to investigate its usefulness in comparison with TV-based regularisation methods through simulation studies using a digital brain phantom. In the present method, the 3D DIP was incorporated into the constrained optimization problem for image denoising as a simultaneous spatial and temporal regularizer, and the optimization problem was solved using the alternating direction method of multipliers (ADMM) (Boyd *et al*. 2011).

## 2. Materials and methods

### 2.1. Denoising method using 3D DIP and ADMM

Image denoising aims to estimate the underlying source image (*f*) from the image corrupted by noise (*g*) under certain constraints. When using the DIP as a constraint tool, this can be reduced to solving the following constrained optimization problem:

$$\hat{\boldsymbol{f}} = \text{argmin}_f \frac{1}{2}\|\boldsymbol{f} - \boldsymbol{g}\|_F^2 \text{ subject to } \boldsymbol{f} = f(\boldsymbol{\theta}|z), \tag{1}$$

where $\hat{\boldsymbol{f}}$ and $\|\cdot\|_F$ denote the denoised image matrix and the Frobenius norm, respectively. The term $f(\boldsymbol{\theta}|z)$ represents the neural network with trainable variables $\theta$ and network input for training z (Ulyanov *et al*. 2018). In this case, the augmented Lagrangian function (Afonso *et al*. 2011) is expressed as

$$\mathcal{L}(\boldsymbol{f}, \boldsymbol{g}, \boldsymbol{\theta}, \boldsymbol{\lambda}, \mu) = \frac{1}{2}\|\boldsymbol{f} - \boldsymbol{g}\|_F^2 + \langle \boldsymbol{\lambda}, \boldsymbol{f} - f(\boldsymbol{\theta}|z)\rangle + \frac{\mu}{2}\|\boldsymbol{f} - f(\boldsymbol{\theta}|z)\|_F^2, \tag{2}$$

where $\boldsymbol{\lambda}$ and $\mu$ ($\geq 0$) denote the Lagrangian multiplier and penalty parameter, respectively, and $\langle \cdot, \cdot \rangle$ denotes the trace inner product.

When applying the ADMM to equation (2), the image matrix ($\boldsymbol{f}_k$), trainable variables ($\boldsymbol{\theta}_k$), and multiplier ($\boldsymbol{\lambda}_k$) at each iteration *k* are computed iteratively as follows (Boyd *et al*. 2011):

$$\boldsymbol{f}_{k+1} = \underset{\boldsymbol{f}}{\text{argmin}} \frac{1}{2}\|\boldsymbol{f} - \boldsymbol{g}\|_F^2 + \langle \boldsymbol{\lambda}_k, \boldsymbol{f} - f(\boldsymbol{\theta}_k|z)\rangle + \frac{\mu}{2}\|\boldsymbol{f} - f(\boldsymbol{\theta}_k|z)\|_F^2 \ (k = 1,2,\cdots, N_{DIP}) \tag{3}$$

$$\boldsymbol{\theta}_{k+1} = \underset{\boldsymbol{\theta}}{\text{argmin}} \ \langle \boldsymbol{\lambda}_k, \boldsymbol{f}_{k+1} - f(\boldsymbol{\theta}|z)\rangle + \frac{\mu}{2}\|\mathbf{f}_{k+1} - f(\boldsymbol{\theta}|z)\|_F^2 \ (k = 1,2,\cdots, N_{DIP}) \tag{4}$$

and

$$\boldsymbol{\lambda}_{k+1} = \boldsymbol{\lambda}_k + \mu[\boldsymbol{f}_{k+1} - f(\boldsymbol{\theta}_{k+1}|z)] \ (k = 1,2,\cdots, N_{DIP}), \tag{5}$$

where $N_{DIP}$ is the number of iterations. When introducing a new parameter ($\boldsymbol{\beta}_k$) defined by $\boldsymbol{\beta}_k = \boldsymbol{\lambda}_k/\mu$ and rearranging equations (3) to (5), the following equations are obtained:

$$\boldsymbol{f}_{k+1} = \underset{\boldsymbol{f}}{\mathrm{argmin}}\ \frac{1}{2}\|\boldsymbol{f}-\boldsymbol{g}\|_F^2 + \frac{\mu}{2}\|\boldsymbol{f}-f(\boldsymbol{\theta}_k|z)+\boldsymbol{\beta}_k\|_F^2 \quad (k=1,2,\cdots,N_{DIP}) \quad (6)$$

$$\boldsymbol{\theta}_{k+1} = \underset{\boldsymbol{\theta}}{\mathrm{argmin}}\ \frac{\mu}{2}\|\boldsymbol{f}_{k+1}-f(\boldsymbol{\theta}|z)+\boldsymbol{\beta}_k\|_F^2 \quad (k=1,2,\cdots,N_{DIP}) \quad (7)$$

and

$$\boldsymbol{\beta}_{k+1} = \boldsymbol{\beta}_k + [\boldsymbol{f}_{k+1}-f(\boldsymbol{\theta}_{k+1}|z)] \quad (k=1,2,\cdots,N_{DIP}). \quad (8)$$

Note that when deriving equations (6) and (7), constant terms are neglected.

Differentiating the right-hand side of equation (6) with respect to $\boldsymbol{f}$ and subsequently setting it to zero, we obtain $\boldsymbol{f}_{k+1}$ as follows:

$$\boldsymbol{f}_{k+1} = \frac{\boldsymbol{g}+\mu[f(\boldsymbol{\theta}_k|z)-\boldsymbol{\beta}_k]}{1+\mu} \quad (k=1,2,\cdots,N_{DIP}). \quad (9)$$

In this study, when calculating $\boldsymbol{\theta}_k$ by training the 3D DIP at each iteration, $k$ (equation (7)), the iteration number for training was fixed at 1000 to avoid overtraining. When calculating $\boldsymbol{f}$ using the ADMM described above, the initial estimates of $\boldsymbol{f}$ ($\boldsymbol{f}_1$) and $\boldsymbol{\beta}$ ($\boldsymbol{\beta}_1$) were set to zero matrices, and $N_{DIP}$ and $\mu$ were fixed at 10 and 0.5, respectively. The above denoising method using the 3D DIP is referred to as "DIP method".

## 2.2. 3D DIP

For solving equation (7), the DIP proposed by Ulyanov *et al*. (2018) and extended to 3D is used. In the DIP method, the network input z in $f(\boldsymbol{\theta}|z)$ is given by random noise; thus, no training pairs are required (Ulyanov *et al*. 2018). The network structure used in this study is illustrated in figure 1, which is an extension of the original two-dimensional (2D) symmetric encoder-decoder architecture (Ulyanov *et al*. 2018) with some modifications (Gong et al. 2019). As illustrated in figure 1, the encoding blocks consisted of repetitive applications of two 3×3×3 3D convolutional layers, each followed by batch normalization (BN) and a leaky rectified linear unit (LReLU). For downsampling, a 3×3×3 3D convolutional layer with a stride of two, followed by BN and LReLU was used. The decoding blocks also consisted of repetitive applications of two 3×3×3 3D convolutional layers, each followed by BN and LReLU, and upsampling and concatenation. For upsampling, 2×2×2 bilinear interpolation was used. The network was optimized by minimizing the loss function (mean squared error) using the Adam optimizer (Kingma and Ba 2014) with a learning rate of $10^{-2}$. The standard deviation (SD) of the random noise used as the network input was fixed at 1/30 according to the original 2D DIP (Ulyanov *et al.* 2018). The neural network was implemented in Python 3.7.12 using Keras on Tensorflow 2.3.0 backend (github.com/keras.team/keras). Training and optimization were performed using an NVIDIA Tesla T4 graphics processing unit (GPU) on the Google Cloud Platform. When the DIP method was used, the computation time was approximately 90 min.

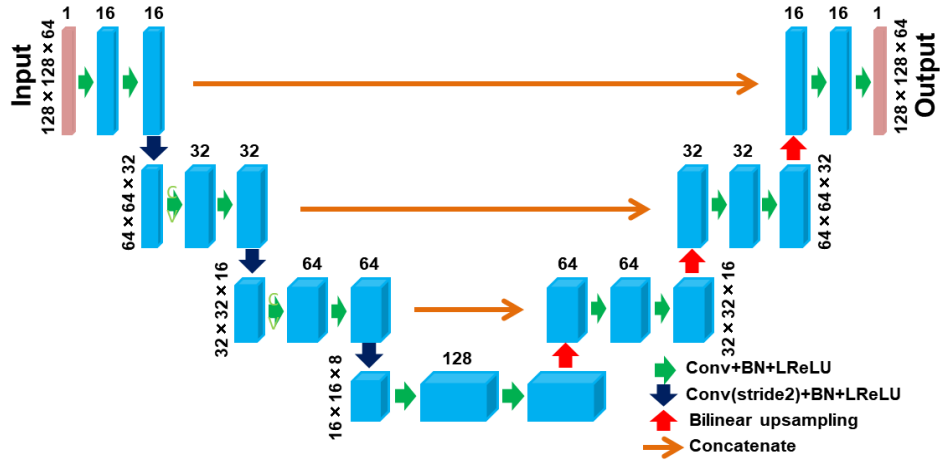

**Figure 1.** Illustration of three-dimensional deep image prior (3D DIP) architecture. Blue boxes represent feature maps. The number on each feature map denotes the number of channels. Conv, BN, and LReLU denote convolution, batch normalization, and leaky rectified linear unit, respectively.

*2.3. Denoising method using TV and ADMM*

When using the TV as a regularizer, the constrained optimization problem to be solved is expressed as

$$\hat{\boldsymbol{f}} = \mathrm{argmin}_{\boldsymbol{f}} \frac{1}{2}\|\boldsymbol{f}-\boldsymbol{g}\|_F^2 + \alpha\|\boldsymbol{w}\|_{TV} \text{ subject to } \boldsymbol{w} = \nabla\boldsymbol{f}, \qquad (10)$$

where $\alpha$ ($\geq 0$) and $\|\cdot\|_{TV}$ denote the regularization parameter and the TV norm, respectively. In this study, we used anisotropic TV, and thus, $\|\cdot\|_{TV}$ denotes the sum of the absolute values of each element ($\ell_1$ norm). The symbol $\nabla$ denotes the 3D spatial and temporal gradient operator ($=(\nabla_x, \nabla_y, \nabla_t)$). In this case, the augmented Lagrangian function (Afonso *et al.* 2011) is expressed as

$$\mathcal{L}(\boldsymbol{f},\boldsymbol{g},\boldsymbol{w},\alpha,\boldsymbol{\lambda},\mu) = \frac{1}{2}\|\boldsymbol{f}-\boldsymbol{g}\|_F^2 + \alpha\|\boldsymbol{w}\|_{TV} + \langle\boldsymbol{\lambda},\nabla\boldsymbol{f}-\boldsymbol{w}\rangle + \frac{\mu}{2}\|\nabla\boldsymbol{f}-\boldsymbol{w}\|_F^2. \qquad (11)$$

As in equations (6)–(8), when applying the ADMM to equation (11) and introducing $\boldsymbol{\beta}_k(=\boldsymbol{\lambda}_k/\mu)$, the image matrix ($\boldsymbol{f}_k$) and multipliers ($\boldsymbol{\beta}_k$ and $\boldsymbol{w}_k$) at each iteration $k$ are computed iteratively, as follows:

$$\boldsymbol{f}_{k+1} = \underset{\boldsymbol{f}}{\operatorname{argmin}} \frac{1}{2}\|\boldsymbol{f} - \boldsymbol{g}\|_F^2 + \frac{\mu}{2}\|\nabla \boldsymbol{f} - \boldsymbol{w}_k + \boldsymbol{\beta}_k\|_F^2 \quad (k = 1,2,\cdots, N_{TV}) \tag{12}$$

$$\boldsymbol{w}_{k+1} = \underset{\boldsymbol{w}}{\operatorname{argmin}} \frac{\mu}{2}\|\nabla \boldsymbol{f}_{k+1} + \boldsymbol{\beta}_k - \boldsymbol{w}\|_F^2 + \alpha\|\boldsymbol{w}\|_{TV} \quad (k = 1,2,\cdots, N_{TY}) \tag{13}$$

and

$$\boldsymbol{\beta}_{k+1} = \boldsymbol{\beta}_k + (\nabla \boldsymbol{f}_{k+1} - \boldsymbol{w}_{k+1}) \quad (k = 1,2,\cdots, N_{TY}), \tag{14}$$

where $N_{TV}$ is the number of iterations. As in equation (9), when differentiating the right-hand side of equation (12) with respect to $\boldsymbol{f}$ and setting it to zero, we obtain the following equation for $\boldsymbol{f}_{k+1}$:

$$(\mathbf{I} - \mu \cdot \Delta)\boldsymbol{f}_{k+1} = \boldsymbol{g} - \mu \cdot \operatorname{div}(\boldsymbol{w}_k - \boldsymbol{\beta}_k) \quad (k = 1,2,\cdots, N_{TY}), \tag{15}$$

where $\mathbf{I}$ denotes an identity matrix. The symbols $\Delta$ and div denote Laplacian and divergence operators, respectively. In equation (15), $\boldsymbol{f}_{k+1}$ cannot be obtained as a closed-form solution, unlike in equation (9). Thus, in this study, we calculated $\boldsymbol{f}_{k+1}$ by solving equation (15) using the conjugate-gradient algorithm (Press *et al.* 1992): For $\boldsymbol{w}_{k+1}$, it can be obtained from equation (13) as follows:

$$\boldsymbol{w}_{k+1} = S_{\alpha/\mu}(\nabla \boldsymbol{f}_{k+1} + \boldsymbol{\beta}_k) \quad (k = 1,2,\cdots, N_{TY}), \tag{16}$$

where $S_{\alpha/\mu}(x)$ is the soft thresholding (shrinkage) operator defined element-wise as $S_{\alpha/\mu}(x) = \operatorname{sgn}(x) \cdot \max(|x| - \alpha/\mu, 0)$.

Starting with the initial estimates ($\boldsymbol{f}_1$, $\boldsymbol{\beta}_1$, and $\boldsymbol{w}_1$) set to zero matrices, the above iterative procedure was repeated until $\|\boldsymbol{f}_{k+1} - \boldsymbol{f}_k\|_F/\|\boldsymbol{f}_k\|_F < \varepsilon_{tol}$ was satisfied or $N_{TV}$ reached 1000. In this study, $\varepsilon_{tol}$ was set to $10^{-6}$, and $\mu$ was set to 0.5. The $\alpha$ values were 0.001, 0.005, and 0.01, respectively. The above TV-based regularization method is referred to as "TV method" and the TV methods with $\alpha$ values of 0.001, 0.005, and 0.01 are denoted by TV(0.001), TV(0.005), and TV(0.01), respectively. When the TV method was used, the GPU was not used, and the computation time was approximately 20 min on average.

## 2.4. Simulation study using DCE-CT images

In this study, a realistic digital brain phantom with bones and blood vessels developed by Aichert *et al*. (2013) was used to simulate DCE-CT cerebral perfusion studies. The DCE-CT images for each slice were generated with the assumption that the images comprised 128×128 pixels in size and 64 frames with a sampling time of 0.8 s. In this study, a relatively low spatial resolution was used to address computational efficiency concerns.

To simulate statistical noise, Gaussian noise was added to DCE-CT images generated using the X-ray exposure noise model (Fang *et al*. 2013). When using this model, the noise SD ($\sigma_n$) in Hounsfield unit (HU) is expressed using the exposure $E$ (mAs) (product of the X-ray tube current and exposure time) as follows (Fang *et al*. 2013):

$$\sigma_n = \sqrt{\sigma_0^2 + \sigma_a^2}, \tag{17}$$

where

$$\sigma_0^2 = \frac{K^2}{E_0} \tag{18}$$

and

$$\sigma_a^2 = \frac{K^2(E_0 - E)}{E_0 E} \tag{19}$$

with $K$ = 103.09 mAs$^{1/2}$ and $E_0$ = 190 mAs. In this study, *E* was varied at 15, 30, 50, 75, and 100 mAs. The $\sigma_n$ values for these exposures are 26.6, 18.8, 14.6, 11.9, and 10.3 HU, respectively.

Before applying the DIP and TV methods, the intensities of the DCE-CT images were scaled to a range between 0 and 1. The DCE-CT images without noise are referred to as "Ground truth (GT) images", whereas those with noise but not processed by the DIP or TV method are referred to as "Original images".

## 2.5. Generation of CBF images

According to the indicator dilution theory for intravascular contrast agents, the time-dependent concentration of the contrast agent, that is, the contrast enhancement (CE) curve in the volume of interest (VOI) at time *t* ($C_{VOI}(t)$), is expressed by (Remp *et al*. 1994)

$$C_{VOI}(t) = \frac{\rho}{k_H} CBF \int_0^t C_{AIF}(\tau) \cdot R(t - \tau) d\tau. \tag{20}$$

In equation (20), $C_{AIF}(t)$ represents the CE curve of the feeding artery. $R(t)$ is the residual function, which is the relative amount of contrast agent in the VOI in an idealized perfusion experiment, where a unit area bolus is instantaneously injected ($R(0) = 1$) and subsequently washed out by perfusion ($R(\infty) = 0$). The symbols $\rho$ and $k_H$ denote the density of brain tissue (1.04 g/mL) and the correction factor for considering the difference between the haematocrit in large and small vessels (0.73), respectively (Remp *et al*. 1994).

From equation (20), the initial height of the deconvolved CE curve equals the CBF multiplied by $\rho/k_H$. In this study, we adopted an algebraic approach based on block-circulant truncated singular value decomposition (bSVD) for deconvolution, which is robust against statistical noise and insensitive to arrival time delay (Wu *et al*. 2003), and calculated the CBF value (mL/100g/min) from the maximum of the deconvolved CE curve. When using bSVD for deconvolution, the elements in the diagonal matrix obtained by bSVD were set to zero when they were smaller than the threshold value given beforehand. In this study, the threshold value was set to 0.2 (Murase *et al*. 2001). We generated CBF images pixel by pixel by applying this approach. In this study, the arterial input function ($C_{AIF}(t)$ in equation (20)) was obtained by manually drawing a region of interest (ROI) on the internal carotid artery (ICA) in the DCE-CT image containing the ICA (slice no. 93 in the digital brain phantom (Aichert *et al*. 2013)).

In this study, CE images were generatited by calculating the average of each pixel intensity of the DCE-CT images between the 5th and 10th frames (baseline) and subtracting them from each pixel intensity. The images from the 5th to 60th frames were used to generate the CBF images.

*2.6. Evaluation*

The original CE images and those obtained using the DIP and TV methods were quantitatively evaluated using two measures. The first measure was the peak signal-to-noise ratio (P*SNR*), defined as

$$PSNR = -10 \cdot \log_{10} \frac{\|\hat{f}_{CE} - \hat{f}_{GT}\|_F^2}{N_{tot} \cdot \max(\hat{f}_{GT})^2}, \tag{21}$$

where $\hat{f}_{CE}$ and $\hat{f}_{GT}$ denote the CE image matrix obtained by the DIP or TV method and the GT image matrix, respectively, $N_{tot}$ denotes the total number of pixels, and $\max(\hat{f}_{GT})$ denotes the maximum value of $\hat{f}_{GT}$.

The second measure is the structural similarity index measure (SSIM) proposed by Wang *et al.* (2004), which is based on the hypothesis that the human visual system is highly adapted to extract structural information. The SSIM for each frame was calculated as follows:

$$SSIM = \frac{(2\mu_{CE}\mu_{GT}+C_1)(2\sigma_{CG}+C_2)}{(\mu_{CE}^2+\mu_{GT}^2+C_1)(\sigma_{CE}^2+\sigma_{GT}^2+C_2)}, \qquad (22)$$

where $\mu_{CE}$ and $\mu_{GT}$ denote the averages within the window set on the original CE image or that obtained by the DIP or TV method and GT image, respectively; $\sigma_{CE}^2$ and $\sigma_{GT}^2$ denote the variances within the window set on the CE and GT images, respectively; $\sigma_{CG}$ denotes the covariance between the window set on the CE and GT images. The terms $C_1$ and $C_2$ denote two variables that stabilize the division with a small denominator. In this study, we calculated the SSIM for each frame using equation (22) with the window size and $C_1$ and $C_2$ set to the default values given by Wang *et al.* (2004) and used the average for all frames to evaluate the overall structural similarity between the original CE images or those obtained by the DIP or TV method and GT images.

*2.7. Linear regression analysis*

Linear regression analysis was performed to quantitatively compare the CBF images generated from the CE images obtained using different methods and those generated from the GT images (figure 4(a)) pixel by pixel. The correlation coefficient between the CBF values obtained from the GT images (x) and those obtained by different methods (y), and the slope and y-intercept of the regression equation between them were calculated.

*2.8. Statistical analysis*

PSNR and SSIM were calculated using equations (21) and (22), respectively. These calculations were performed five times for each method, and the means and SDs of these measures for *n* = 5 were calculated. When linear regression analysis was performed, CBF images were also generated five times for each method, and the correlation coefficient, slope, and y-intercept of the regression equation and their mean and SD values for *n* = 5 were calculated.

One-way analysis of variance was used to compare the above measures and parameters among groups. As a post-hoc test, the Tukey-Kramer multiple comparison test was used to determine statistical significance. A *p* value less than 0.05 was considered statistically significant.

## 3. Results

Figure 2(a) shows the typical CE curves obtained from the GT images (GT) (red solid line), original CE images (original) (blue solid line), and CE images obtained by the DIP method (DIP) (green solid line) at an exposure of 15 mAs, whereas figure 2(b) shows those for GT (red solid line), TV(0.001) (blue solid line), TV(0.005) (green solid line), and TV(0.01) (black solid line). These curves represent the temporal changes in the mean CE value (HU) within the square ROI with 10×10 pixels drawn on the right occipital region (black dotted line in figure 4(a)). When using the DIP method, the noise was significantly reduced, and the curve was closest to that of the GT images (figure 2(a)). When using TV(0.001) and TV(0.005), noise reduction was insufficient. For TV(0.01), the noise reduction was remarkable, but the curve was excessively smooth (figure 2(b)).

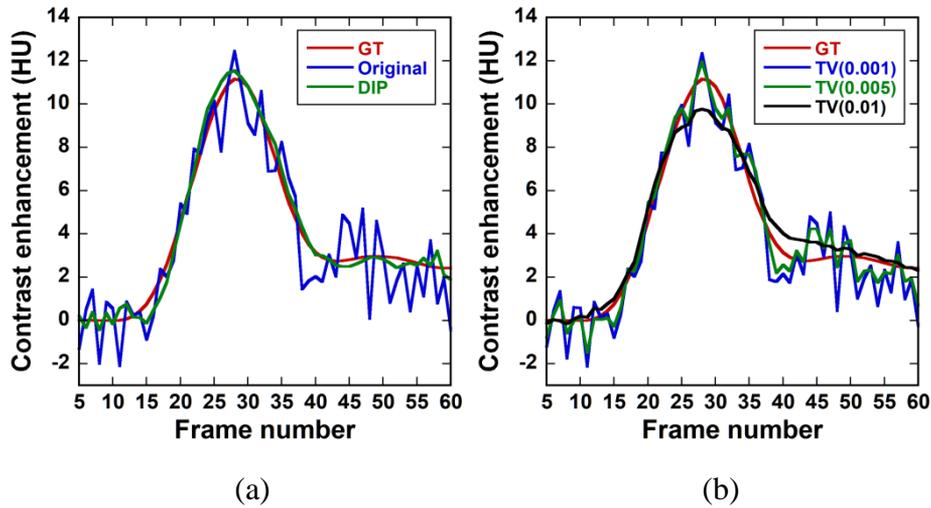

(a)                                      (b)

**Figure 2.** (a) Contrast enhancement (CE) as a function of frame number at an exposure (product of X-ray tube current and exposure time) of 15 mAs. The red, blue, and green solid lines show the CE curves obtained from the ground truth (GT) images, original CE images (Original), and CE images obtained by the DIP method (DIP), respectively. (b) CE as a function of frame number at an exposure of 15 mAs. The red, blue, green, and black solid lines show the CE curves obtained from the GT images and CE images obtained by the total variation (TV)-based method with regularization parameter ($\alpha$) values of 0.001 (TV(0.001)), 0.005 (TV(0.005)), and 0.01 (TV(0.01)), respectively. These curves represent the temporal changes in the mean CE value in Hounsfield unit (HU) within the square region of interest (ROI) with 100 pixels, as indicated by the black dotted line in figure 4(a).

Figure 3(a) shows the PSNR values for the original (red solid line), DIP (blue solid line), TV(0.001) (green solid line), TV(0.005) (green dashed line), and TV(0.01) (green dotted line) as a function of exposure, whereas their SSIM values are shown in figure 3(b). As shown in figure 3, the PSNR and SSIM values for DIP did not change significantly depending on the exposure, whereas those for the original and TV methods were significantly dependent on the exposure. Furthermore, when using the TV methods, they significantly changed depending on the $\alpha$ value. When the exposure was 15 mAs, there were significant differences in PSNR other than between the original and TV(0.001) and between DIP and TV(0.01). There were significant differences in SSIM other than between DIP and TV(0.01). When the exposure was 30 mAs, there were significant differences in both PSNR and SSIM other than between the original and TV(0.001), between DIP and TV(0.005), between DIP and TV(0.01), and between TV(0.005) and TV(0.01). When the exposure was 50 mAs, there were significant differences in PSNR other than between the original and TV(0.001), between the original and TV(0.01), and between TV(0.001) and TV(0.01). There were significant differences in SSIM other than between the original and TV(0.001), between the original and TV(0.01), and between DIP and TV(0.005). When the exposure was 75 mAs, there were significant differences in both PSNR and SSIM, except between DIP and TV(0.005). When the exposure was 100 mAs, significant differences were observed in PSNR other than between DIP and TV(0.005) and between TV(0.001) and TV(0.005). For SSIM, significant differences were observed except between DIP and TV(0.005).

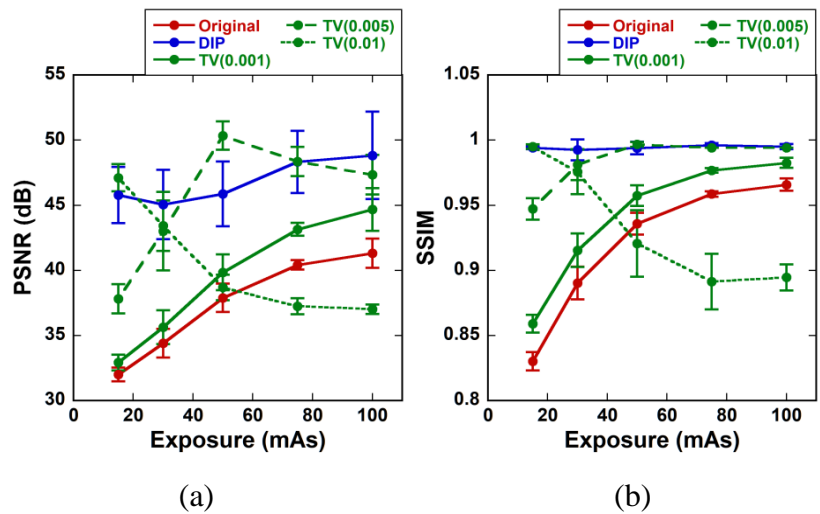

(a)            (b)

**Figure 3.** (a) Relationship between peak signal-to-noise ratio (PSNR) (equation (21)) and exposure for different methods. (b) Relationship between structural similarity index measure (SSIM) (equation (22)) and exposure for different methods. The red solid, blue solid, green solid, green dashed, and green dotted lines show the original, DIP, TV(0.001), TV(0.005), and TV(0.01), respectively. Data are represented as mean ± standard deviation (SD) for $n = 5$.

Figure 4(a) shows the CBF image generated from the GT images at the level of the caudate nucleus (slice no. 140 in the digital brain phantom (Aichert *et al*. 2013)), whereas figure 4(b) shows the typical CBF images generated from the original CE images and those obtained by DIP, TV(0.001), TV(0.005), and TV(0.01) (from left to right columns) for different exposures. As shown in figure 4, when using the DIP method, the noise was reduced even at 15 mAs. When using the TV method, the noise decreased with increasing $\alpha$ value, whereas the spatial resolution deteriorated.

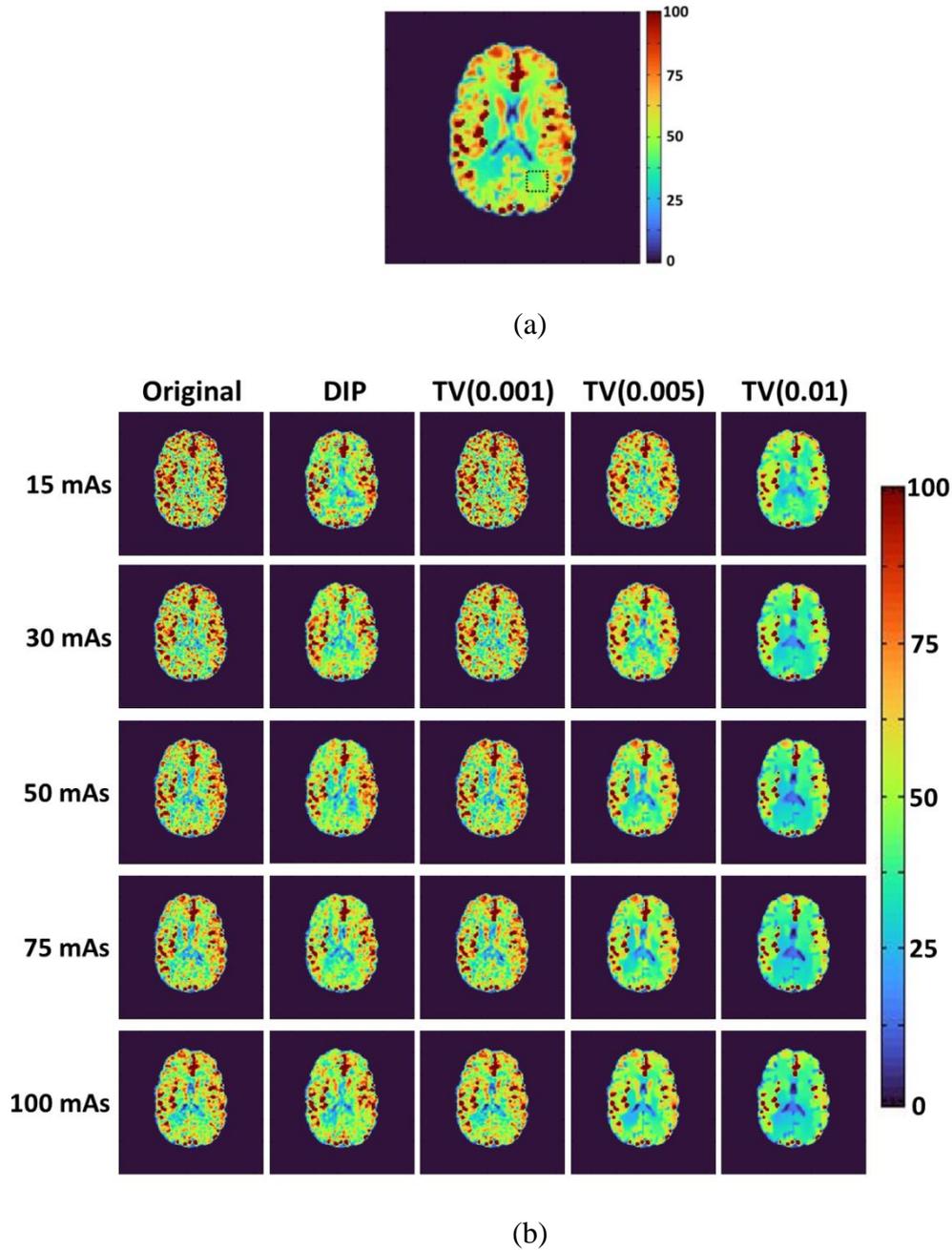

(a)

(b)

**Figure 4.** (a) Cerebral blood flow (CBF) image generated from the GT images at the level of the caudate nucleus (slice no. 140 in the digital brain phantom (Aichert *et al*. 2013)). The black dotted square shows the ROI for calculating the CE curves shown in figure 2. (b) CBF images generated from the original CE images and the CE images obtained by DIP, TV(0.001), TV(0.005), and TV(0.01) (from left to right columns) at exposures of 15 mAs (first row), 30 mAs (second row), 50 mAs (third row), 75 mAs (fourth row), and 100 mAs (fifth row). The CBF images are displayed with the same window ranging from 0 to 100 mL/100g/min.

The results of the linear regression analysis are shown in figure 5. Figures 5(a), 5(b), and 5(c) show the correlation coefficient, slope, and y-intercept of the regression equation as a function of exposure, respectively. The red solid, blue solid, green solid, green dashed, and green dotted lines represent the original, DIP, TV(0.001), TV(0.005), and TV(0.01), respectively. When the exposure was 15 mAs, there were significant differences in the correlation coefficients other than between the original and TV(0.001) and between DIP and TV(0.005). There were significant differences in the slope between the original and DIP, between DIP and TV(0.001), between DIP and TV(0.005), between DIP and TV(0.01), and in the y-intercept other than between the original and TV(0.001). When the exposure was 30 mAs, significant differences were observed in the correlation coefficients other than between TV(0.005) and TV(0.01). There were significant differences in the slope other than between the original and TV(0.001) and in the y-intercept for all combinations of methods. When the exposure was 50 mAs, there were significant differences in the correlation coefficients other than between the original and TV(0.001) and between DIP and TV(0.001). As in the case with an exposure of 30 mAs, there were significant differences in the slope other than between the original and TV(0.001) and in the y-intercept for all combinations of methods. When the exposure was 75 mAs, there were significant differences in the correlation coefficients other than between TV(0.001) and TV(0.01), and in the slope other than between the original and TV(0.001), between the original and TV(0.005), and between TV(0.001) and TV(0.005). For the y-intercept, significant differences were observed for all combinations of methods. When the exposure was 100 mAs, there were significant differences in the correlation coefficients other than between the original and TV(0.001), between the original and TV(0.01), and between TV(0.001) and TV(0.01). As in the case with an exposure of 75 mAs, significant differences were observed in the slope other than between the original and TV(0.001), between the original and TV(0.005), and between TV(0.001) and TV(0.005), and in the y-intercept for all combinations of methods.

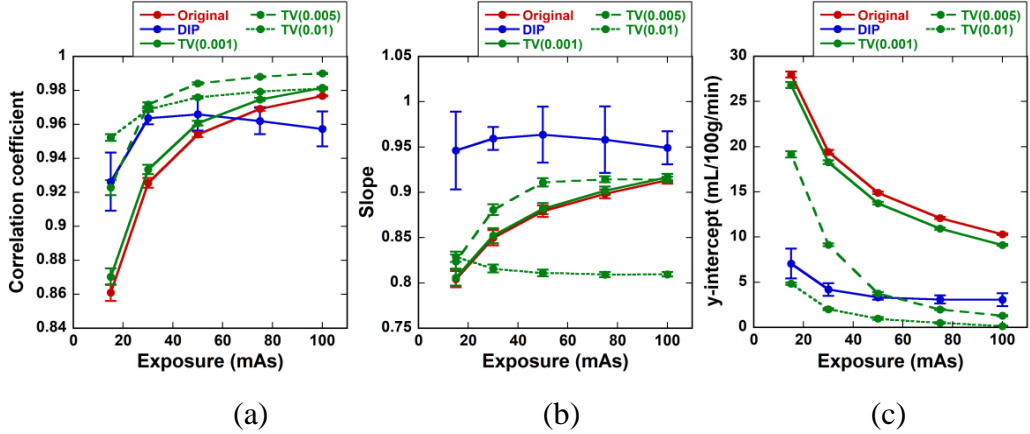

|   (a)   |   (b)   |   (c)   |

**Figure 5.** Results of linear regression analysis between the CBF image generated from the GT images (x) and that generated from the CE images obtained by different methods (y). (a) Relationship between the correlation coefficient and exposure. (b) Relationship between the slope of regression equation and exposure. (c) Relationship between the y-intercept of regression equation and exposure. The red solid, blue solid, green solid, green dashed, and green dotted lines show the original, DIP, TV(0.001), TV(0.005), and TV(0.01), respectively. Data are represented as mean ± SD for $n = 5$.

## 4. Discussion

In this study, we presented a denoising method based on 3D DIP and applied it to DCE-CT brain perfusion studies. We also investigated its usefulness in comparison with TV methods through simulation studies using a digital brain phantom. In the DIP method, the constrained optimization problem given by equation (1) was solved using ADMM, in which 3D DIP was used as a simultaneous spatial and temporal regularizer. Our preliminary results (figures 2–5) suggest that the DIP method is useful for denoising DCE-CT images at ultra-low to low exposures ($\leq$ ~50 mAs).

When 2D DIP was applied to the 2D DCE-CT image at each time frame, a different CNN for each time frame was trained; thus, the continuity of the image intensity among time frames was lost, resulting in an increase in bias among time frames (data not shown). Therefore, when applying 2D DIP to 3D DCE-CT images, it is necessary to add a bias compensation term to the loss function used in training the CNN, as done by Wu *et al.* (2021).

In this study, we extended 2D DIP (Ulyanov *et al.* 2018) to 3D DIP using 3D convolutional layers (figure 1). When 3D DIP alone was applied to 3D DCE-CT images without any constraints, the noise was reduced in the images obtained by 3D DIP alone, but the image intensities largely deviated from those of the GT images (data not shown). As proposed in this study, when 3D DIP is incorporated into the constrained optimization problem for image denoising (equation (1)) as a simultaneous spatial and temporal regularizer, such deviation is reduced (figure 2). Thus, it appears that the proposed DIP method exhibits both denoising and bias compensation effects.

When applying the TV method, it is important to search for optimal values of the hyperparameters. The tuning parameters in this study were $\alpha$ in equation (10) and $\mu$ in equation (11). Finding these optimal values is not a straightforward task and is performed by trial and error using hand tuning, which requires considerable time and effort to find them from many candidates. In the TV method, the regularization parameter $\alpha$ controls the sparsity (piece-wise smoothness) of the processed images through TV norm minimization. The sparsity (piece-wise smoothness) increases with increasing $\alpha$ value, and vice versa. As shown in figure 2, when $\alpha$ is small, the noise reduction is insufficient, whereas the CE curve became blurred due to excessive smoothing when $\alpha$ was taken large. This finding is also confirmed by the fact that the PSNR and SSIM for the TV method were significantly dependent on the $\alpha$ value (figure 3) and by the visual inspection of the CBF images (figure 4(b)). Furthermore, the TV norm minimization causes staircase artefacts. When $\alpha$ was 0.005 or 0.01, these artefacts were observed (figure 2).

However, the DIP method has no hyperparameters that need to be tuned, except for $\mu$ in equation (2). As shown in figure 3, the PSNR and SSIM for the DIP method were not significantly dependent on the exposure, that is, the noise level, contrary to those in the TV method. Thus, it appears that the DIP method is more adaptable and flexible than conventional TV-based regularization methods and is useful for reducing the time and effort required for parameter tuning. The fact that no parameter tuning is required appears to be important, especially for practical applications, because the GT images required for parameter tuning are unknown in practical applications.

The penalty parameter $\mu$ in equations (2) and (11) is associated with the convergence behavior and stability in solving these equations because this parameter determines the step size in the iterative computation of the ADMM (Murase 2021). In this study, this parameter was empirically chosen to be 0.5 in both the DIP and TV methods, considering both the convergence speed and stability of the ADMM.

There were significant differences between the DIP and TV methods in linear regression analysis (figure 5). When the original CE images or TV methods were used, the correlation coefficient between the CBF values obtained by these methods and those obtained from the GT images increased with increasing exposure. However, when using the DIP method, a peak was observed at an exposure of 50 mAs, which decreased gradually thereafter (figure 5(a)). As previously described, the SD of the random noise and iteration number for training the CNNs in the DIP method were fixed to constants (1/30 and 1000, respectively) for simplicity throughout the studies. Thus, the above finding may be because the overtraining effect increased or the noise was overestimated compared to the actual noise level with increasing exposure (decreasing noise level) and it might be necessary to change the SD and iteration number according to the noise level of the input images. In contrast, the slope of the regression equation for the DIP method was closer to unity than those of the other methods at all the exposures studied (figure 5(b)), and the y-intercept (offset) was significantly smaller than those for the other methods, except for TV(0.01), at an exposure ≤ 50 mAs (figure 5(c)), suggesting that the linearity of the CBF values obtained by the DIP method is superior to those obtained by the other methods. This may reflect the fact that the SSIM for the DIP method was the highest for all the exposures studied (figure 3). These features of the DIP method appear to be promising, especially when diagnosing and/or predicting the outcome of patients with cerebrovascular diseases, because the DIP method allows noise reduction in CBF images without reducing their linearity.

The drawback of the present DIP method is that the SDs of the measures, such as PSNR, correlation coefficient, and slope, are larger than those obtained from the original CE images and by the TV methods (figures 3 and 5). Therefore, further studies are necessary to improve the stability of the proposed method. As described above, random noise was used as the network input according to the original DIP (Ulyanov *et al*. 2018) in this study, because it is easy to implement. As proposed by Gong *et al*. (2019), using magnetic resonance imaging or CT images of each patient as the network input, instead of random noise, may be effective for stabilizing the present DIP method. These studies are currently underway.

## 5. Conclusion

We presented a denoising method for DCE-CT images for cerebral perfusion studies using 3D DIP and ADMM, and investigated its usefulness in comparison with TV-based regularization methods through simulation studies. Our preliminary results suggest that the present DIP method is useful for denoising DCE-CT images at ultra-low to low exposures (≤ ~50 mAs) and for improving the accuracy of the CBF images generated from them, although further studies using a wide range of clinical data and comparative studies with other existing methods are necessary to establish the present method.

**Acknowledgement**

The author is grateful to Prof. Noriyuki Tomiyama and Dr. Nobuo Kashiwagi for giving him an opportunity to perform the present study.

**Conflicts of interest**

The author declares no conflicts of interest.

**References**

Afonso M V, Bioucas-Dias J M and Figueiredo M A T 2011 An augmented Lagrangian approach to the constrained optimization formulation of imaging inverse problems *IEEE Trans. Image Process.* **20** 681−695.

Aichert A, Manhart M T, Navalpakkam B K, Grimm R, Hutter J, Maier A, Hornegger J and Doerfler A 2013 A realistic digital phantom for perfusion C-arm CT based on MRI data. *Proc. IEEE Nucl. Sci. Symp. Med. Imag. Conf. Rec. (NSS/MIC)* Seoul, South Korea, pp. 1–2.

Beck A and Tebouble M 2009 Fast gradient-based algorithms for constrained total variation image denoising and deblurring problems *IEEE Trans. Image Process.* **18** 2419–2434.

Boyd S, Parikh N, Chu E, Peleato B and Eckstein J 2011 Distributed optimization and statistical learning via the alternating direction method of multipliers *Found. Trends Mach. Learn.* **3** 1–122.

Chen H, Zhang Y, Zhang W, Liao P, Li K, Zhou J and Wang G 2017 Low-dose CT via convolutional neural network *Biomed. Opt. Express* **8** 679–694.

Fang R, Chen T and Sanelli P C 2013 Towards robust deconvolution of low-dose perfusion CT: sparse perfusion deconvolution using online dictionary learning *Med. Image Anal.* **17** 417–428.

Gong K, Catana C, Qi J and Li Q 2019 PET image reconstruction using deep image prior *IEEE Trans. Med. Imaging* **38** 1655–1665.

Hirata M, Sugawara Y, Fukutomo Y, Oomoto K, Murase K and Mochizuki T 2004 Measurement of radiation dose in cerebral CT perfusion study *Rad. Med.* **23** 97–103.

Kadimesetty V S, Gutta S, Ganapathy S and Yalavarthy P K 2019 Convolutional neural network-based robust denoising of low-dose computed tomography perfusion maps *IEEE Trans. Radiat. Plasma Med. Sci.* **3** 137–152.

Kingma D and Ba J 2014 Adam: a method for stochastic optimization arXiv:1412:6980.

Li S, Zeng D, Peng J, Bian Z, Zhang H, Xie Q, Wang Y, Liao Y, Zhang S, Huang J, Meng D,

Xu Z and Ma J 2019 An efficient iterative cerebral perfusion CT reconstruction via low-rank tensor decomposition with spatial–temporal total variation regularization *IEEE Trans. Med. Imaging* **38** 360–370.

Lustig M, Donoho D and Pauly J M. 2007 Sparse MRI: the application of compressed sensing for rapid MR imaging *Magn. Reson. Med.* **58** 1182–1195.

Mendrik A M, Vonken E J, Van Ginneken B, De Jong H W, Riordan A, Van Seeters T, Smit E J, Viergever M A and Prokop M 2011 TIPS bilateral noise reduction in 4D CT perfusion scans produces high-quality cerebral blood flow maps *Phys. Med. Biol.* **56** 3857–3872.

Miles K A 1991 Measurement of tissue perfusion by dynamic computed tomography *Br. J. Radiol.* **64** 409–412.

Murase K, Shinohara M and Yamazaki Y 2001 Accuracy of deconvolution analysis based on singular value decomposition for quantification of cerebral blood flow using dynamic susceptibility contrast-enhanced magnetic resonance imaging *Phys. Med. Biol.* **46** 3147–3159.

Murase K, Nanjo T, Ii S, Miyazaki S, Hirata M, Sugawara Y, Kudo M, Sasaki K and Mochizuki T 2005 Effect of x-ray tube current on the accuracy of cerebral perfusion parameters obtained by CT perfusion studies *Phys. Med. Biol.* **50** 5019–5029.

Murase K, Nanjo T, Sugawara Y, Hirata M and Mochizuki T 2015 Usefulness of an anisotropic diffusion method in cerebral CT perfusion study using multi-detector row CT *Open J. Med. Imaging* **5** 106-116.

Murase K 2021 Alternating direction method of multipliers applied to medical image restoration arXiv:2107.01653.

Nagayoshi M, Murase K, Fujino K, Uenishi Y, Kawamata M, Nakamura Y, Kitamura K, Higuchi I, Oku N and Hatazawa J 2005 Usefulness of noise adaptive non-linear Gaussian filter in FDG-PET study *Ann. Nucl. Med.* **19** 469–477.

Niu S, Zhang S, Huang J, Bian Z, Chen W, Yu G, Liang Z and Ma J 2016 Low-dose cerebral perfusion computed tomography image restoration via low-rank and total variation regularizations *Neurocomputing* **197** 143–160.

Press W H, Teukolsky S A, Vetterling W T and Flannery B P 1992 Numerical Recipes in C. Oxford: Cambridge University Press.

Remp K A, Brix G, Wenz F, Becker C R, Gückel F and Lorenz W J 1994 Quantification of regional cerebral blood flow and volume with dynamic susceptibility contrast-enhanced MR imaging *Radiology* **193** 637–641.

Ulyanov D, Vedaldi A and Lempitsky V 2018 Deep image prior *Proc. IEEE Conf. Comput. Vis. Pattern Recognit.* pp. 9446–9454.

Wang Z, Bovik A C, Sheikh H R and Simoncelli E P 2004 Image quality assessment: from error visibility to structural similarity *IEEE Trans. Image Process.* **13** 600–612.

Wintermark M, Reichart M, Thiran J-P, Maeder P, Chalaron M, Schnyder P, Bogousslavsky J and Meuli R 2002 Prognostic accuracy of cerebral blood flow measurement by perfusion computed tomography, at the time of emergency room admission, in acute stroke patients. *Ann. Neurol.* **51** 417–432.

Wu O, Østergaard L, Weisskoff R M, Benner T, Rosen B R and Sorensen A G 2003 Tracer arrival time-insensitive technique for estimating flow in MR perfusion-weighted imaging using singular value decomposition with a block-circulant deconvolution matrix *Magn. Reson. Med.* **50** 164–174.

Wu D, Ren H and Li Q 2021 Self-supervised dynamic CT perfusion image denoising with deep neural networks *IEEE Trans. Radiat. Plasma Med. Sci.* **5** 350–361.

Yoo J, Jin K H, Gupta H, Yerly J, Stuber M and Unser M 2021 Time-dependent deep image prior for dynamic MRI *IEEE Trans. Med. Imaging* **40** 3337 –3348.

Zeng D, Zhang X, Bian Z, Huang J, Zhang H, Lu L, Lyu W, Zhang J, Feng Q, Chen W and Ma J 2016 Cerebral perfusion computed tomography deconvolution via structure tensor total variation regularization *Med. Phys.* **43** 2091–2107.

Zhang K, Zuo W, Chen Y, Meng D and Zhang L 2017 Beyond a Gaussian denoiser: residual learning of deep CNN for image denoising *IEEE Trans. Image Process.* **26** 3142–3155.

Zhou K C and Horstmeyer R 2020 Diffraction tomography with a deep image prior. *Opt. Express* **28** 12872–12896.